\documentclass[dvips,12pt,a4paper]{article}
\usepackage{epsfig}
\usepackage{rotate}
\usepackage{a4wide}
\usepackage{amssymb}

\def\lsim{\mathrel{\rlap{\raise 2.5pt \hbox{$<$}}\lower 2.5pt
\hbox{$\sim$}}}
\def\air{\vphantom{\bigg|}}

\newlength{\absize}
\begin{document}
\thispagestyle{empty}


\begin{flushright}
{\tt University of Bergen, Department of Physics}    \\[4pt]
{\tt Scientific/Technical Report No.~2000-03}    \\[4pt]
{\tt ISSN 0803-2696} \\[5pt]
{hep-ph/0006343} \\[2pt]
{June 2000}       
\end{flushright}
\vspace*{2.0cm}

\begin{center}
{\bf \Large Vacuum and MSW interpretations of solar neutrino data 
            with the LNW mass matrix}
\end{center}
\vspace{0.5cm}

\begin{center}
Per Osland and Geir Vigdel
\end{center}
\vspace{1cm}

\begin{center}
Department of Physics, University of Bergen,
     Allegt.\ 55, N-5007 Bergen, Norway
\end{center}
\vspace*{1.5cm}

\begin{abstract}
The Lehmann--Newton--Wu mass matrix, which was recently applied
to neutrinos, is further investigated. The analytic results presented
earlier are confirmed numerically for the solar
density profile of the Standard Solar Model.
The combined analysis of atmospheric and solar neutrino data
favors the MSW solution over the vacuum-oscillation solution.
The total rates from the solar neutrino detectors favor 
two regions in the ($m_1, m_2, m_3$) mass space,
with the heaviest one, $m_3\sim 0.05$~eV.
The spectrum distortion reported by the Super-Kamiokande collaboration 
favors one of these two regions,
which has two light ($m_1, m_2\lsim 0.003$~eV) neutrinos.
\end{abstract}

\vfill
\section{Introduction}
\setcounter{equation}{0}
Neutrino flavor transitions (e.g.~$\nu_e \rightarrow \nu_\mu$) are the most
widely accepted explanation of the atmospheric neutrino anomaly and
the solar neutrino problem (SNP) \cite{review}.  
In the most popular scenarios for such flavor conversions, 
vacuum oscillations and matter enhanced transitions (MSW) \cite{MSW}, 
the neutrino conversion and survival probabilities 
depend on two types of neutrino-specific parameters.  
One of them is the mixing matrix $U$ defined from 
$\nu_\alpha = \sum U_{\alpha i} \nu_i$, 
where $\alpha = e, \mu, \mbox{ or } \tau$ and the
sum runs over the mass states, $\nu_1$, $\nu_2$, $\nu_3$.  
The other is the differences of masses squared, 
$\Delta m_{kj}^2 = m_k^2 - m_j^2$, where the latter
refer to the neutrino mass states $\nu_i$.  
There is no widely accepted mass matrix,
therefore the conventional way of analyzing the results from neutrino
oscillation experiments is to consider the mixing matrix as independent of
the mass-squared differences.  
But when we assume a certain mass matrix $M$ in the Lagrangian
(in a current basis)
\begin{equation}\label{Lagr}
{\cal L} = \bar{\nu} M \nu + {\rm h.c.},
\end{equation}
then the mixing will explicitly depend on the masses through the
diagonalization of $M$.
The maximal mixing arising from the democratic texture \cite{FrXi} is a
widely discussed example of this kind (see also \cite{Altarelli}).

The mass matrix proposed by Lehmann, Newton and Wu (LNW) 
for quarks \cite{LNW}, 
was recently applied to neutrinos in \cite{OW},
and gross features of the model were explored
for solar and atmospheric neutrinos.
We shall here examine the implications of that model in more detail,
focusing on solar neutrinos, and compare the vacuum-oscillation scenario
with the MSW  scenario.
Also, we will for the different cases discuss the energy spectrum 
of the recoil electrons induced by solar neutrinos.
Constraints from atmospheric neutrino data will be taken into account.

\section{The model}
\setcounter{equation}{0}
Let us briefly review the model being considered \cite{LNW,OW}.
One assumes that there are three neutrinos 
with a real mass matrix of the form:
\begin{equation}
M=\left(\begin{array}{ccc}
0 & d & 0 \\ d & c & b\\ 0 & b & a
\end{array}\right),
\end{equation}
where $b^2=8c^2$.
This symmetric mass matrix is diagonalized by the matrix $U$, 
\begin{equation}
\label{Eq:def-U}
M_d = U^\dagger MU,
\end{equation}
where $M_d$ is a diagonal matrix whose entries are the mass eigenvalues.  
The diagonalization involves solving a cubic equation for $a$.

For an arbitrary set of mass eigenvalues, ($m_1$, $m_2$, $m_3$),
the model yields two possible mixing matrices (two physical solutions
of the cubic equation), referred to as Solution~1 and Solution~2 \cite{OW}.
The corresponding mixing matrices $U$ are displayed in 
Fig.~\ref{sol1-2} for $m_2\le m_3$
(the physical region of the model extends a little beyond the triangle 
given by $m_1\le m_2 \le m_3$).  
As illustrated in those figures, the mixing elements are determined by
the two mass ratios $m_1/m_3$ and $m_2/m_3$.  
\begin{figure}[htb]
\refstepcounter{figure}
\label{sol1-2}
\addtocounter{figure}{-1}
\begin{center}
\setlength{\unitlength}{1cm}
\begin{picture}(9.0,8.3)
\put(-4.3,0.0)
{\mbox{\epsfysize=8.8cm\epsffile{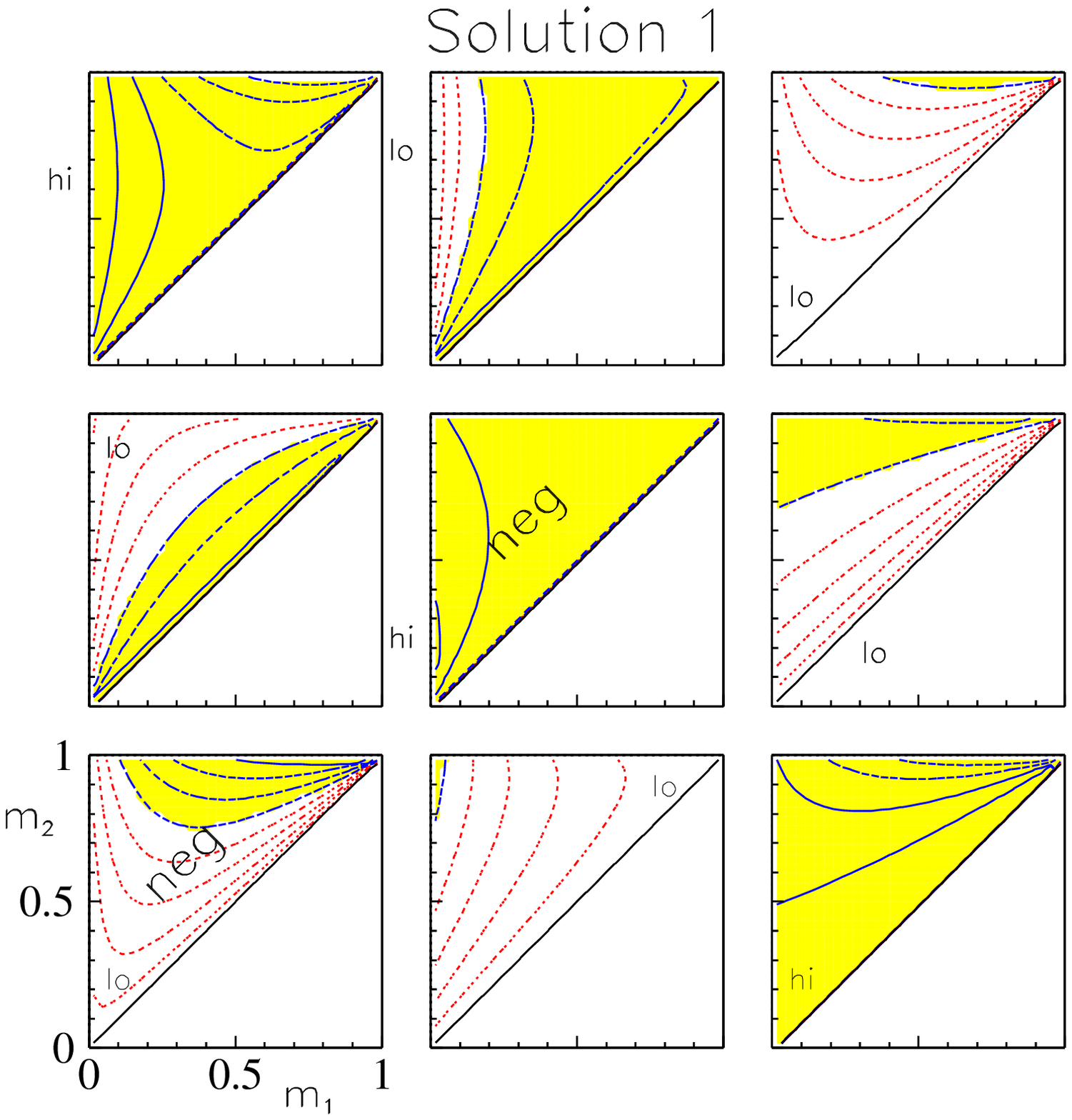}}
 \mbox{\epsfysize=8.8cm\epsffile{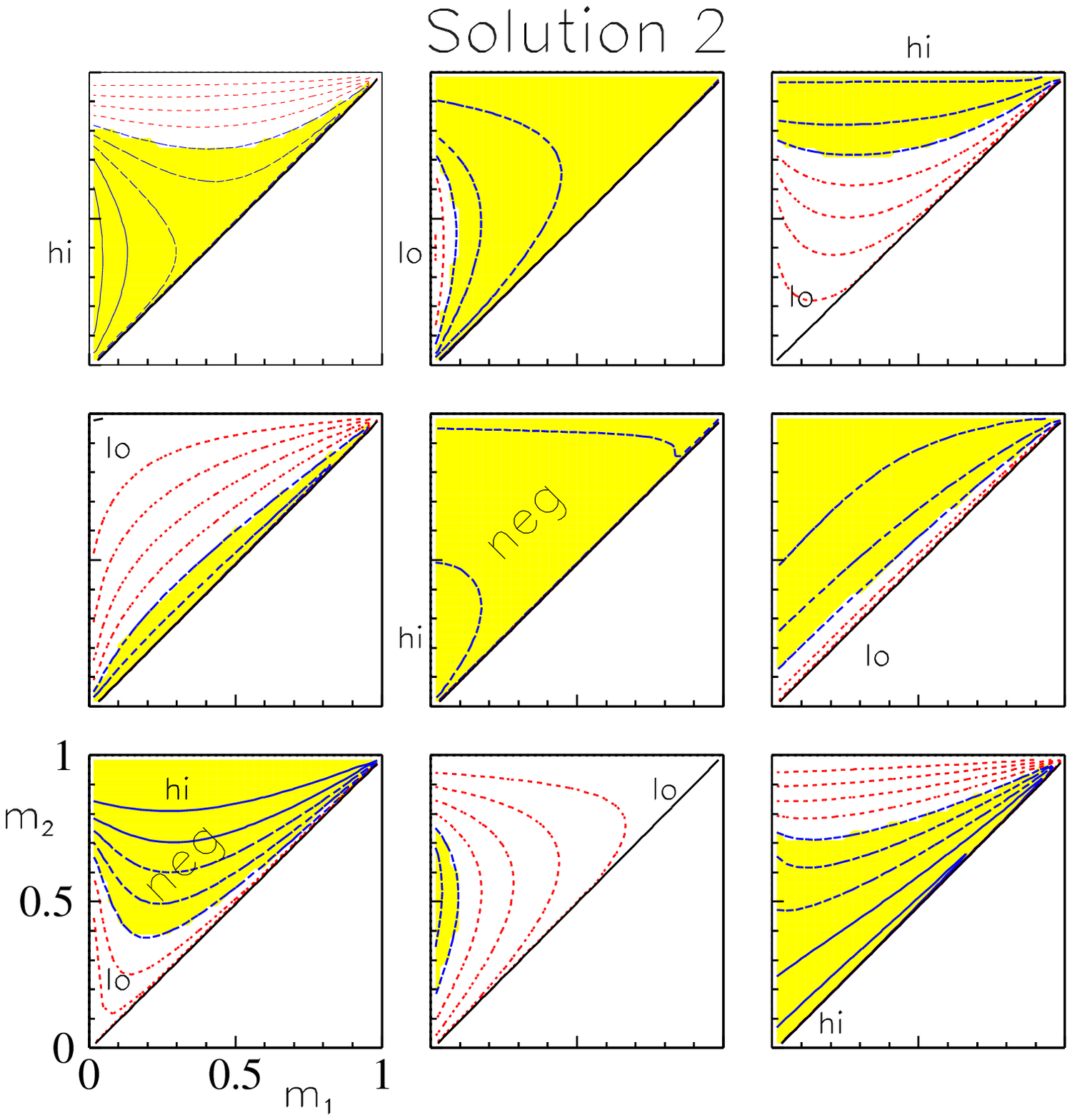}}}
\end{picture}
\caption{Absolute values of mixing matrix elements,
$U_{e1}$, $U_{e2}$, $U_{e3}$, \ldots
{\it vs.}\ $m_1$ and $m_2$
($m_3=1$) for Solutions~1 and 2.
Dotted contours: 0.1, 0.2, 0.3, 0.4; dashed: 0.5, 0.6, 0.7;
solid: 0.8, 0.9 (regions above 0.5 are shaded).
Phases are chosen such that the angles in Eq.~(\ref{Eq:PDG-matrix})
are positive.
Elements $(2,2)$ and $(3,1)$ are negative.}
\end{center}
\end{figure}

The neutrino oscillation data give quite strong restrictions on both 
the mixing and the mass-squared differences, when treated as free parameters.
Also, a model for the neutrino masses should not violate the experimental
upper bounds on the neutrino masses \cite{PDG98}: $m_{\nu_e}\lsim 10$~eV, 
$m_{\nu_\mu}<0.17$~MeV, $m_{\nu_\tau} < 18.2$~MeV.  
Furthermore, it should satisfy the mass constraints from cosmology 
\cite{cosmology}: $\sum m_{\nu_f} \lsim 94$~eV,
be capable of explaining the apparent oscillation of atmospheric neutrinos, 
with $\Delta m^2 \sim 10^{-3}~\mbox{eV}^2$, 
and give a plausible explanation of the SNP, 
with $\Delta m^2 \sim 10^{-10}~\mbox{eV}^2$ (vacuum oscillation)
or $\Delta m^2 \sim 10^{-5}~\mbox{eV}^2$ (MSW).  
The natural scale implied by
these values for the mass-squared differences is far too small to 
be in conflict with any of the above-mentioned mass bounds.  

We denote the mass states in such a way that $m_1 \le m_3$.  Of the two
independent $\Delta m_{kj}^2$, one will be reserved for an explanation of the
atmospheric neutrino anomaly, and the other will be assigned to a solution of
the solar neutrino problem.  Thus we are trying in {\it one} model to satisfy 
the two most serious neutrino problems.  
According to our notation, the smallest mass-squared difference will be 
either $\Delta m_{21}^2$ or $\Delta m_{32}^2$.

In the present model one can fit directly in terms of {\em masses} 
as opposed to mass-squared differences and mixing factors.
When the values of the mixing elements are given by the masses, we
do not expect as good a theoretical fit to the observed data as when these
parameters are considered as independent.  
For given values of $U$ and $\Delta m_{kj}^2$ the probability 
expression for
flavor conversion will be the same whether the mixing matrix is dependent on
the masses or not.  Therefore we know to some extent from the traditional
analyses where in the parameter space we can expect acceptable fits
for the model we are analyzing.  For example, for solar neutrinos
it is known that both vacuum- and matter-influenced neutrino transitions
require small $U_{e3}$ in order to give good fits. 
From Fig.~\ref{sol1-2} we see that this is achieved 
when both $m_1/m_3$ and $m_2/m_3$ are small.  
And from the same figure it is apparent that in this region
the mixing elements $U_{e1}$ and $U_{e2}$ are very sensitive to small
changes in the masses.  This means that by small changes in
the mass ratios, the mixing elements can be adjusted close to the
best-fit values obtained when $U$ and the masses are considered as
independent.

\begin{table}[htb]
\begin{center}
\begin{tabular}{|c|c|c|c|}
\hline
Detector & Observed Rate & Predicted Rate& Obs./Pred. \\
\hline \hline
\air Homestake  \cite{Homestake}&$2.56\pm0.16\pm0.16$&$7.7^{+1.2}_{-1.0}$
&$0.332\pm0.029$   \\ \hline
\air SAGE \cite{SAGE}  &$67.2^{+7.2\;+3.5}_{-7.0\;-3.0}$ & $129^{+8}_{-6}$
&$0.521^{+0.062}_{-0.059}$ \\ \hline
\air GALLEX \cite{GALLEX} &$77.5 \pm 6.2^{+4.3}_{-4.7}$&$129^{+8}_{-6}$
&$0.601^{+0.058}_{-0.060}$\\ \hline
\air Kamiokande \cite{Kamiokande}  &$2.80\pm0.19\pm0.33$& $5.15^{+1.0}_{-0.7}$
&$0.544\pm0.074$ \\ \hline
\air Super-K \cite{Super-K-spectrum}  &$2.44\pm0.05^{+0.09}_{-0.07}$
&$5.15^{+1.0}_{-0.7}$&$0.474^{+0.020}_{-0.017}$ \\ \hline
\end{tabular}
\end{center}
\caption{Measured and predicted (no oscillations) \cite{BBP-98} 
event rates for solar neutrino detectors. 
The flux units for the three upper detectors are given in SNU, 
1 event per second per $10^{36}$ target atoms. For the scattering
detectors the fluxes are rated in units of $10^{6} \mbox{ cm}^{-2}{\rm
s}^{-1}$. The rightmost column shows the fraction between observed and
predicted fluxes.}
\label{tabell}
\end{table}
In the analysis of the solar-neutrino data (summarized in Table~1)
we will see that, for small values of $m_1$ and $m_2$,
the two mixing solutions give approximately the same results.  

In the traditional analyses of the data from neutrino-oscillation experiments
one often expresses the amount of mixing in terms of angles.  
We adopt the following, common parameterization of the mixing matrix
\cite{PDG98}:
\begin{eqnarray}
\label{Eq:PDG-matrix}
U= \left[
\begin {array}{ccc} 
c_{12}\, c_{13} & s_{12}\,c_{13} & s_{13}\\
-c_{12}\,s_{13}\,s_{23}-s_{12} \,c_{23} &
-s_{12}\,s_{13}\,s_{23}+c_{12}\,c_{23} & c_{13}\,s_{23}\\
-c_{12}\,s_{13}\,c_{23}+s_{12}\,s_{23}&
-s_{12}\,s_{13}\,c_{23}-c_{12}\,s_{23}& c_{13}\,c_{23}
\end{array}\right],
\end{eqnarray}
where $c_{12}= \cos\theta_{12}$, $s_{12}= \sin\theta_{12}$, etc.  
This will be particularly useful when we discuss the MSW mechanism.
If $\theta_{13}=0$ we get the two-generation case.

For vacuum oscillations as well as for the MSW scenario,
we shall discuss four cases,
small $\Delta m_{21}^2$, small $\Delta m_{32}^2$, together with
Solutions~1 and 2.

\section{Atmospheric-neutrino constraints}
\setcounter{equation}{0}
We shall not here perform a detailed $\chi^2$ fit to
the atmospheric-neutrino data.
Instead, we will use best-fit values for $U$ and $\Delta m_{kj}^2$ 
from the literature as constraints and compare those with the acceptable 
regions we find for the solar neutrino data.
The experimental data indicate that the ``missing'' muon neutrinos can be
explained by a close to maximal $\nu_\mu \leftrightarrow \nu_\tau$
oscillation \cite{Super-K-prl}, which requires small $U_{e3}$.  
Also, the result from the CHOOZ experiment \cite{CHOOZ},
interpreted in terms of three generations and our choice for 
the $\Delta m_{kj}^2$-values, indicates small $U_{e3}$.  
For masses where our model gives good fits to the solar neutrino data, 
we actually have $U_{e3} \ll 1$.  
This criterion allows us to neglect matter effects for atmospheric
neutrinos \cite{Bilenkii}.

With three neutrino families the oscillation probability between arbitrary
flavor states can be written as
\begin{equation}\label{genprob}
P_{\nu_\alpha \leftrightarrow \nu_\beta}(R)
=\delta_{\alpha \beta}-4\sum_{j<k} 
\left[U_{\alpha j}U_{\beta j}U_{\alpha k}U_{\beta k}\right]
\sin^2\left(\frac{\Delta m_{kj}^2R}{4E}\right), \nonumber 
\end{equation}
where $R$ is the distance traveled since the neutrinos were
created, and $E$ is the neutrino energy.

For the case of atmospheric neutrinos, the factor $R/E$ is quite small
as compared with the case of solar neutrinos.
Thus, the term in Eq.~(\ref{genprob}) with a small $\Delta m_{kj}^2$ 
(henceforth referred to as the SNP value, $\sim 10^{-5}~\mbox{eV}^2$ or 
$\sim 10^{-10}~\mbox{eV}^2$) can be neglected.
When this small SNP term involves $\Delta m_{21}^2$ we get
\begin{equation}\label{atmsmalld21}
P_{\nu_\mu\leftrightarrow \nu_\tau}(R) \simeq
4U_{\mu 3}^2U_{\tau 3}^2 \sin^2\left(\frac{\Delta m_{31}^2R}{4E}\right),
\end{equation}
where we use the approximation $\Delta m_{31}^2 \simeq \Delta m_{32}^2$.
From \cite{Super-K-prl} we know that the best fit between data and 
the oscillation hypothesis occurs at 
$\Delta m_{31}^2 \sim 3.2 \times 10^{-3}~\mbox{eV}^2$.
This means that the lowest possible value of the heaviest mass state is 
$m_3 = {\cal O}(0.05 ~\mbox{eV})$.  

In Fig.~\ref{amplatm} we have plotted the amplitude 
in Eq.~(\ref{atmsmalld21}) for Solutions~1 and 2.  
One should keep in mind that the approximation leading to 
Eq.~(\ref{atmsmalld21}) is only satisfied close to the diagonal, $m_1 = m_2$.  
For Solution~1, an unrealistically large $m_2/m_3$ would be needed to get 
an acceptable $\nu_\mu - \nu_\tau$ mixing.
For Solution~2, we observe that large $\nu_\mu - \nu_\tau$ mixing
could be reached for values of $\Delta m_{21}^2$ which are marginally
compatible with the solar-neutrino constraints.
We also note that the $\Delta m_{21}^2$ which is required
for the MSW effect will allow a larger amplitude for atmospheric-neutrino
transitions than a mass-squared difference relevant for 
the vacuum oscillation solution to the SNP, since the MSW effect requires
a larger $m_2$.
\begin{figure}[hbt]
\refstepcounter{figure}
\label{amplatm}
\addtocounter{figure}{-1}
\hspace*{0.5cm}{\epsfysize=7cm
{\mbox{\epsffile{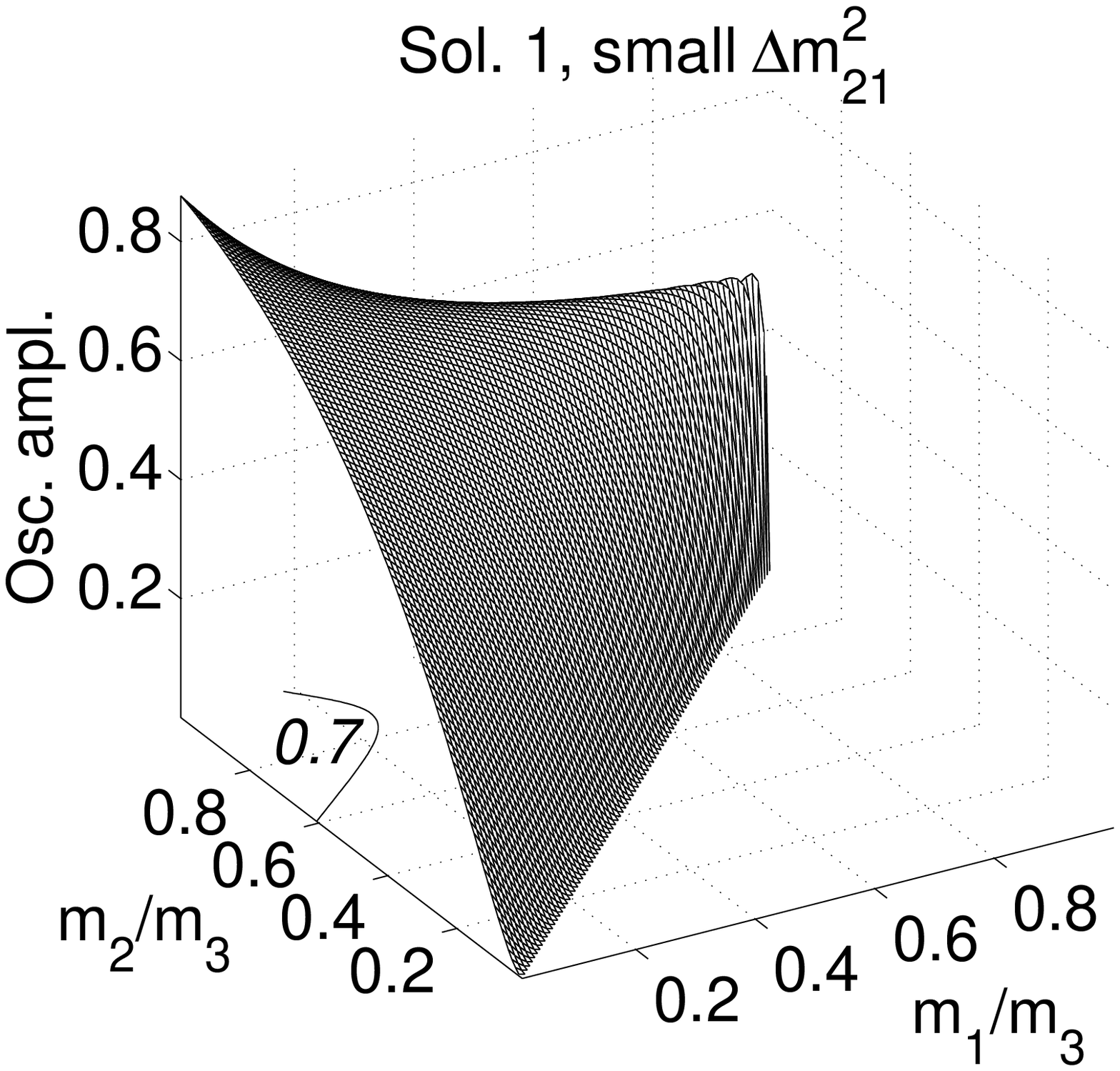}}
 \mbox{\epsffile{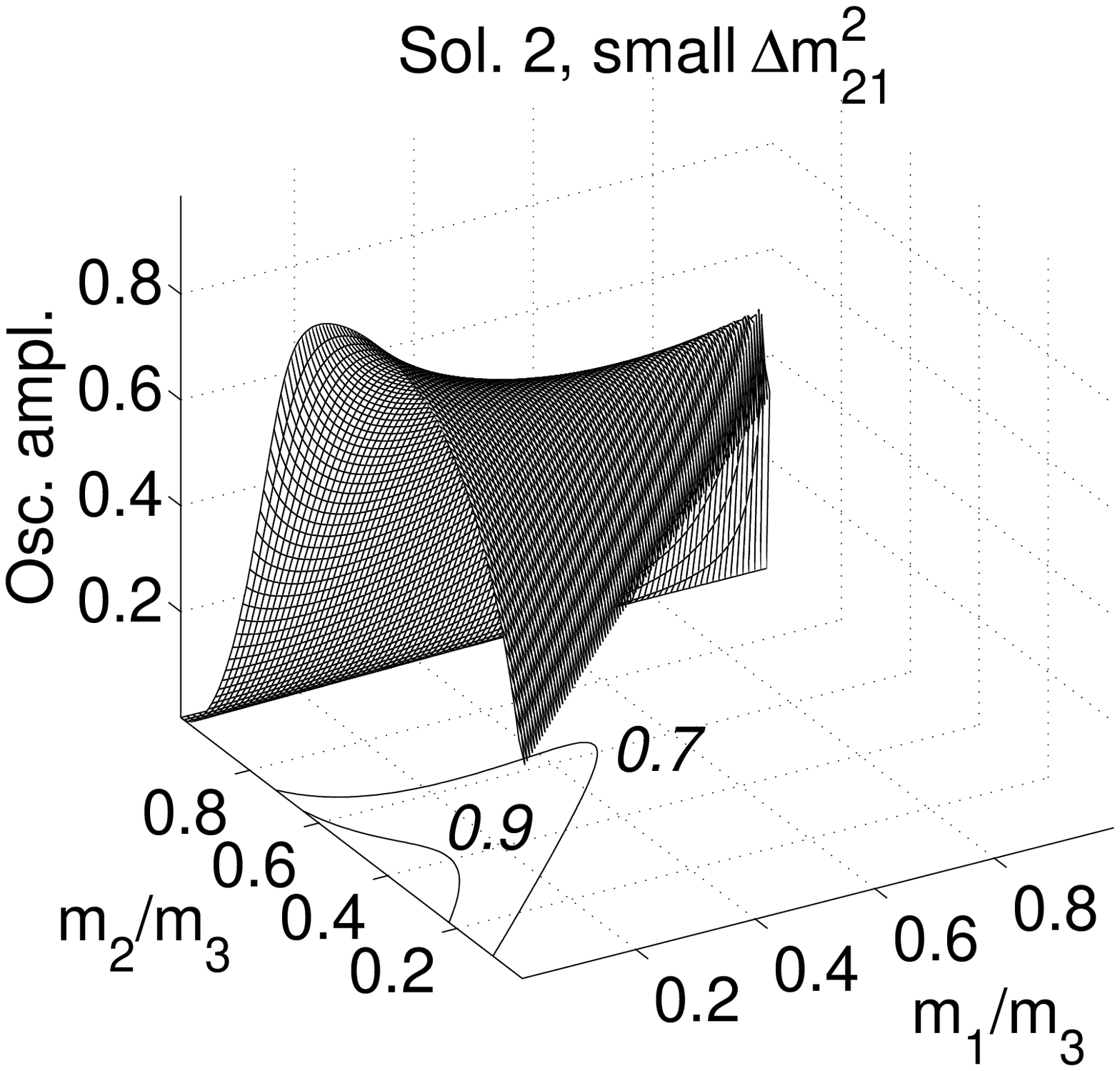}}}}
\caption{Amplitude $4U_{\mu3}^2U_{\tau3}^2$
as function of mass ratios $m_1/m_3$ and $m_2/m_3$,
relevant for atmospheric neutrinos under the assumption
of small $\Delta m_{21}^2$.
The maximal values are around 0.9 and  1 for Solutions~1 and 2, respectively.
Contours corresponding to 0.7 and 0.9 are indicated.}
\end{figure}

When $\Delta m_{32}^2$ assumes the SNP value, i.e., $m_2\simeq m_3$,
then the atmospheric $\nu_\mu\rightarrow \nu_\tau$ transition probability
can be written as $P_{\nu_\mu\leftrightarrow \nu_\tau}(R) 
\simeq 4U_{\mu 1}^2U_{\tau 1}^2 \sin^2\left({\Delta m_{31}^2R}/{4E}\right)$
instead of (\ref{atmsmalld21}),
where we have
used the approximation $\Delta m_{31}^2 \simeq \Delta m_{21}^2$.  The maximal
amplitude for this oscillation is about 0.5 for both mixing solutions. 
This value is too low to be of real interest.

In summary, the atmospheric-neutrino data favor parameters corresponding
to Solution~2 with $\Delta m_{21}^2 \ll \Delta m_{31}^2 
\simeq \Delta m_{32}^2$.
They also favor the MSW solution for the solar-neutrino problem,
since this mechanism requires a larger $m_2$, and thus a larger amplitude
for atmospheric-neutrino oscillations.
\section{Vacuum oscillations}
\setcounter{equation}{0}
Vacuum oscillations provide a plausible solution to the solar-neutrino
problem \cite{OV,Barger}.
In this scenario, the length of the relevant oscillation cycle 
(see Eq.~(\ref{genprob})) is about one Sun-Earth distance, 
i.e. $\sim 1.5 \times 10^{11}$~m.  
The corresponding $\Delta m_{kj}^2$
is several orders of magnitude smaller than the one which is
relevant for atmospheric neutrinos.
As mentioned, it will be either $\Delta m_{21}^2$ or $\Delta m_{32}^2$
which is relevant for solar neutrinos.
Thus, the allowed values of $\Delta m_{32}^2$ will be of the order 
relevant for either atmospheric or solar neutrinos.

One necessary condition for acceptable fit between data and the vacuum
oscillation hypothesis is that the smallest 
$\Delta m_{kj}^2 \sim 10^{-10}~\mbox{eV}^2$. 
The other independent mass-squared difference is held fixed at 
$3.2 \times 10^{-3}~\mbox{eV}^2$, in order to be able to accommodate 
the atmospheric neutrino data.
To find the optimal mixing matrix we have to search through a large 
number of values for the mass ratios $m_1/m_3$ and $m_2/m_3$,
which we do by varying $m_1$ and $\Delta m_{21}^2$.

\subsection{Small \boldmath{$\Delta m_{21}^2$}}
The interesting region is here close to the diagonal in the $m_1$--$m_2$
plane (see Fig.~\ref{sol1-2}).
The best fit is achieved at $m_1=3.1\times 10^{-6}~\mbox{eV}$, 
$\Delta m_{21}^2 = 8.0\times 10^{-11}~\mbox{eV}^2$ which gives $\chi^2=4.1$.
This $\chi^2$-value is about 0.1\% higher than the value obtained in an
analysis where $U$ and $\Delta m_{kj}^2$ are treated as independent. 
This good agreement is explained by the high sensitivity of the mixing
matrix to the mass values.  Fig.~\ref{Ue1Ue2Ue3} shows how the relevant 
mixing elements depend on $m_1$ when $\Delta m_{21}^2$ and
$\Delta m_{32}$ are held fixed at $8.0\times 10^{-11}~\mbox{eV}^2$ and
$3.2\times 10^{-3}~\mbox{eV}^2$, respectively.  As can be seen from this
figure, the mixing solutions are equal for $m_1 \lesssim 10^{-3}~\mbox{eV}$.
For Solution~1, $U_{\mu 3} \ll 1$, which means no 
$\nu_\mu \leftrightarrow \nu_\tau$ for atmospheric neutrinos.  
For Solution~2 this amplitude is $\sim 0.4$.  
This value is unchanged over a broad range of
mass values, including the best fit value for the solar neutrino solution.
In Fig.~\ref{chik} we have plotted the $\chi^2$ values corresponding to
Fig.~\ref{Ue1Ue2Ue3}.  When $m_1$ is reduced beyond the best-fit value, the
mixing matrix approaches the identity matrix, hence the fit becomes poor.

\begin{figure}[htb]
\refstepcounter{figure}
\label{Ue1Ue2Ue3}
\addtocounter{figure}{-1}
\begin{center}
{\epsfysize=8cm{\mbox{\epsffile{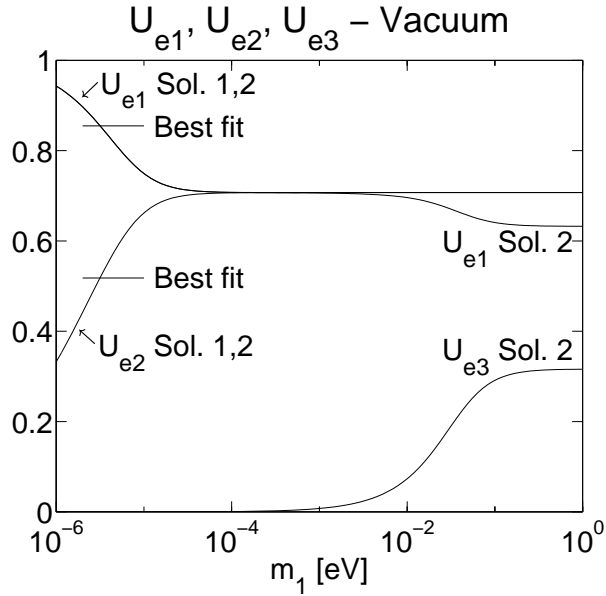}}}}
\end{center}
\caption{Dependences of $U_{e1}$, $U_{e2}$ and $U_{e3}$ on $m_1$. 
The mass-squared differences $\Delta m_{21}^2$ and $\Delta m_{32}^2$
are held fixed at $8.0\times 10^{-11}~\mbox{eV}^2$ and 
$3.2\times 10^{-3}~\mbox{eV}^2$, respectively. 
The two solutions are equal up to $m_1 \simeq 10^{-3}~\mbox{eV}$. 
For Solution~1, the mixing element $U_{e3}$ is close to zero for 
all values of $m_1$.}
\end{figure}

\begin{figure}[htb]
\refstepcounter{figure}
\label{chik}
\addtocounter{figure}{-1}
\begin{center}
{\epsfysize=8cm{\mbox{\epsffile{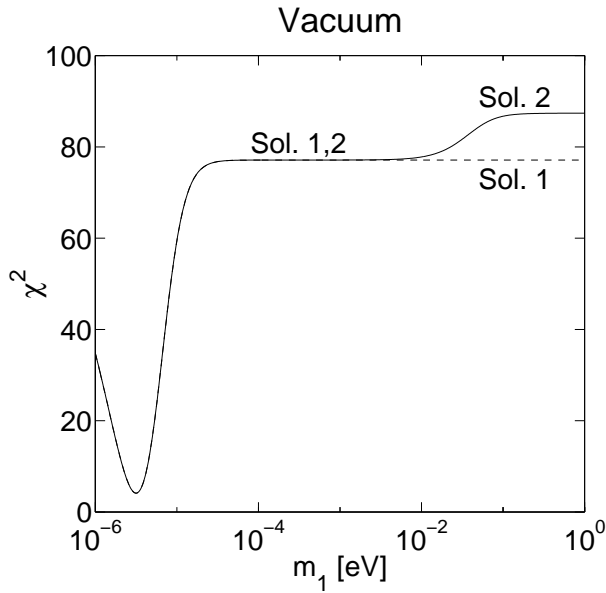}}}}
\end{center}
\caption{Fit to the solar-neutrino data for small $\Delta m_{21}^2$. 
The values for that parameter and for $\Delta m_{32}^2$ are the
same as in Fig.~\ref{Ue1Ue2Ue3}. The best fit gives $\chi^2 = 4.1$ 
at $m_1=3.1\times 10^{-6} ~\mbox{eV}$.}
\end{figure}

\subsection{Small \boldmath{$\Delta m_{32}^2$} ($m_2/m_3 \simeq 1$)}
In this case we can not get an equally good fit for solar neutrinos 
as we had for small $\Delta m_{21}^2$.  
It turns out that the best fit is achieved when we,
in the expression (\ref{genprob}), have vanishing amplitude for 
the oscillation factors containing the largest $\Delta m_{kj}^2$ \cite{OV}.
This can not be achieved within the model, neither for Solution~1
nor for Solution~2, see Fig.~\ref{sol1-2}. 
For Solution~1 the best fit gives $\chi^2 \simeq 6.1$, while
for Solution~2 one gets $U_{e2} \simeq U_{e3} \simeq 1/\sqrt{2}$,
with the best fit result $\chi^2 \simeq 5.0$.
Also, the relevant amplitude for atmospheric neutrino oscillations
would be rather low in this case.

\subsection{Summary on vacuum oscillations}

If we consider only the SNP, none of the best fits can be excluded.
The best-fit point is achieved with small $\Delta m_{21}^2$ 
and at practically the same mass values for Solutions~1 and 2.  
Including atmospheric neutrino data, the preferred solution would be 
small $\Delta m_{21}^2$ and Solution~2.  
But the maximum possible oscillation amplitude
for $\nu_\mu \leftrightarrow \nu_\tau$ is only 0.4, so the vacuum
oscillation solution to the SNP is disfavored by the model in \cite{OW}.
On the other hand, the recoil spectrum has an enhancement at high energies,
as suggested by the data (see Fig.~\ref{spectrum}).
\section{The MSW interpretation}
For an exponential solar density, the three-flavor MSW equations
have recently been solved analytically for arbitrary masses \cite{OW-99}.
Here, since we assume two widely separated values of $\Delta m_{kj}^2$,
we use the more conventional approach \cite{Fogli-Petcov}.
This allows for a numerical comparison of the two approaches.
The agreement was found to be very good.
For example, for ($m_1$, $m_2$, $m_3$) = (0.002, 0.01, 0.057)~eV
and $E=1$~MeV, the two approaches give $\nu_e$ survival probabilities
of 0.668 and 0.660.
For some of the mass points we also check the exponential solar density 
approximation against the solar density profile
calculated in a standard solar model \cite{BBP-98,Bahcall-web}, by performing
numerical integrations on the latter.  The agreement is within a few percent.

In the two-generation MSW case where the angle and mass-squared difference
are varied independently, one arrives at two possible solutions in the
parameter space spanned by $\sin^2(2\theta)$ and $\Delta m^2$.  
There is one region with low $\chi^2$ at small angles, and another
minimum at large angles \cite{indepMSW}.  
As we will see, this is also the case for the model in \cite{OW}.

For solar neutrinos, Solutions~1 and 2 are nearly identical 
for the masses relevant at the best-fit point.  
Therefore  we will divide the presentation of our results into two parts, 
small $\Delta m_{21}^2$ and small $\Delta m_{32}^2$.

\subsection{Small \boldmath{$\Delta m_{21}^2$}}
For small values of $m_2/m_3$ and very small values of $m_1/m_3$ we get
$U_{e1} \simeq 1$ for both mixing solutions.
These are practically equivalent for solar neutrinos, 
both of them have a best fit with $\chi^2 = 1.8$ at 
$m_1=2.8 \times 10^{-6}~\mbox{eV}$, 
$\Delta m_{21}^2=7.0\times 10^{-6}~\mbox{eV}^2$, where we have imposed
$\Delta m_{31}^2 = 3.2 \times 10^{-3}~\mbox{eV}^2$.  
The latter quantity is held fixed at this value, which is compatible with
the atmospheric neutrino data \cite{Super-K-prl}.  Because it is
large compared to the SNP-relevant $\Delta m_{21}^2$, its exact value
is, within the uncertainty, unimportant for the solar-neutrino analysis.
These mass values correspond to small mixing angle, 
see Eq.~(\ref{Eq:PDG-matrix}).
With the above-mentioned values of the mass-squared differences held fixed,
we show in Fig.~\ref{sma-lma} (left panel) how $U_{e1}$ and $\chi^2$ 
vary when $m_1$ is slightly perturbed around its best-fit value.

\begin{figure}[htb]
\refstepcounter{figure}
\label{sma-lma}
\addtocounter{figure}{-1}
\begin{center}
\setlength{\unitlength}{1cm}
\begin{picture}(12,7.0)
\put(-2.0,0.0)
{\mbox{\epsfysize=7.0cm\epsffile{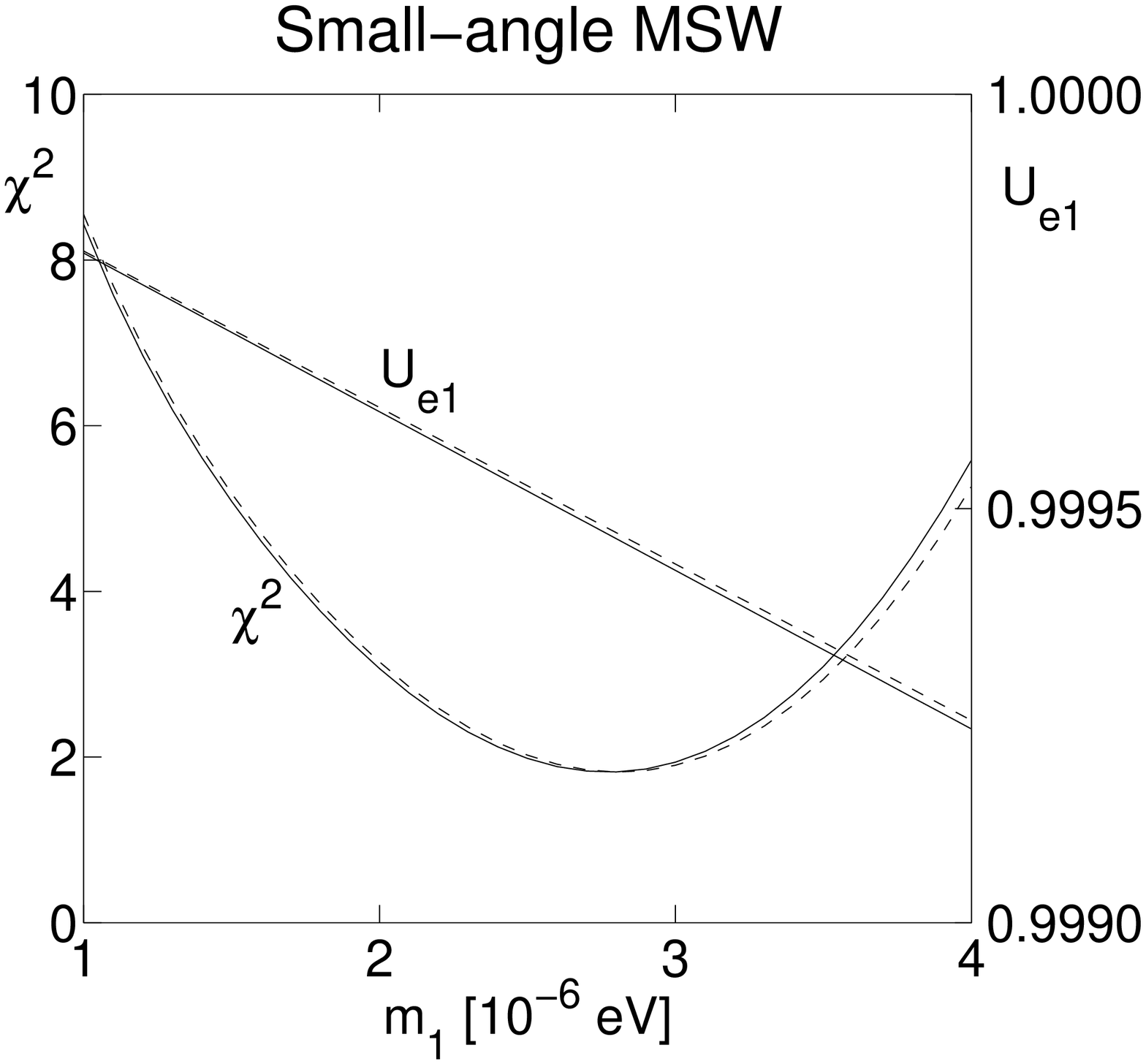}}
 \mbox{\epsfysize=7.0cm\epsffile{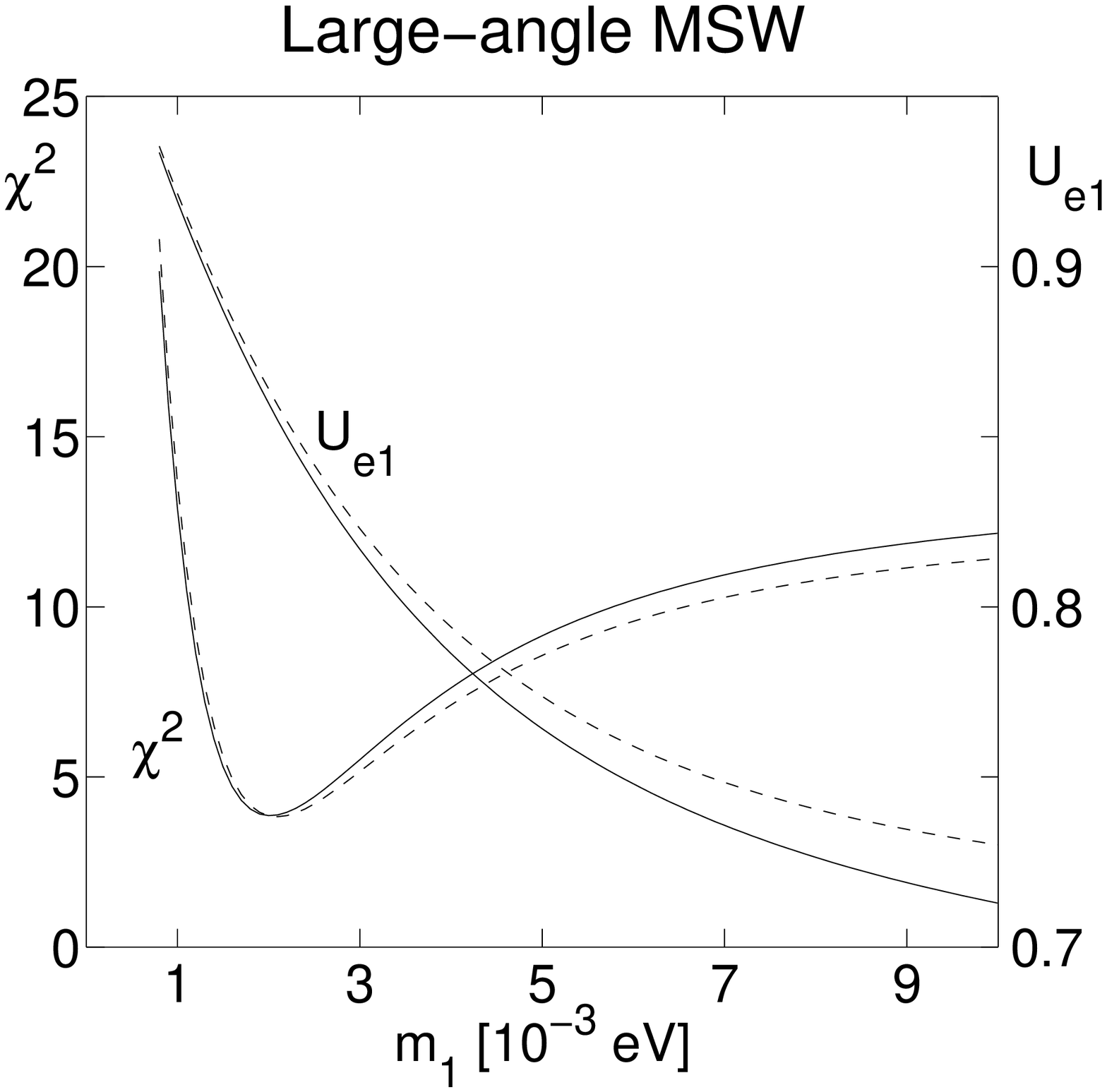}}}
\end{picture}
\vspace*{-3mm}
\end{center}
\caption{Mixing element $U_{e1}$ and $\chi^2$ {\it vs.}\ $m_1$,
for Solution~1 (dashed line) and Solution~2 (solid line) in 
the small-mixing-angle (left panel) and large-mixing-angle (right panel) 
MSW case.
In the SMA case, the best fit has $\chi^2=1.8$ and occurs at 
$m_1=2.8 \times 10^{-6}~\mbox{eV}$, 
$\Delta m_{21}^2= 7.0 \times 10^{-6}~\mbox{eV}^2$.
In the LMA case, the best fit has $\chi^2=3.8$ at $m_1=2.1
\times 10^{-3}~\mbox{eV}$, $\Delta m_{21}^2=3.2 \times 10^{-5}~\mbox{eV}^2$.
In both cases, we have imposed
$\Delta m_{31}^2 = 3.2 \times 10^{-3}~\mbox{eV}^2$.}
\end{figure}

As previously noted, for atmospheric neutrinos
we consider only vacuum oscillations, due to the smallness of $U_{e3}$.
In the parameter region shown in Fig.~\ref{sma-lma} (left panel), 
$U_{e3} < 10^{-3}$.
From Eq.~(\ref{atmsmalld21}) it follows that 
the $\nu_\mu \leftrightarrow \nu_\tau$ oscillation amplitude is 0.05 
and 0.6 for Solutions~1 and 2, respectively. 
Thus, Solution~1 is unacceptable in this parameter region.
We also note that the $\chi^2$ value is quite sensitive 
to the mass values $m_1$ and $m_2$.

\begin{figure}[htb]
\refstepcounter{figure}
\label{sma+lma}
\addtocounter{figure}{-1}
\begin{center}
{\epsfysize=7.5cm
{\mbox{\epsffile{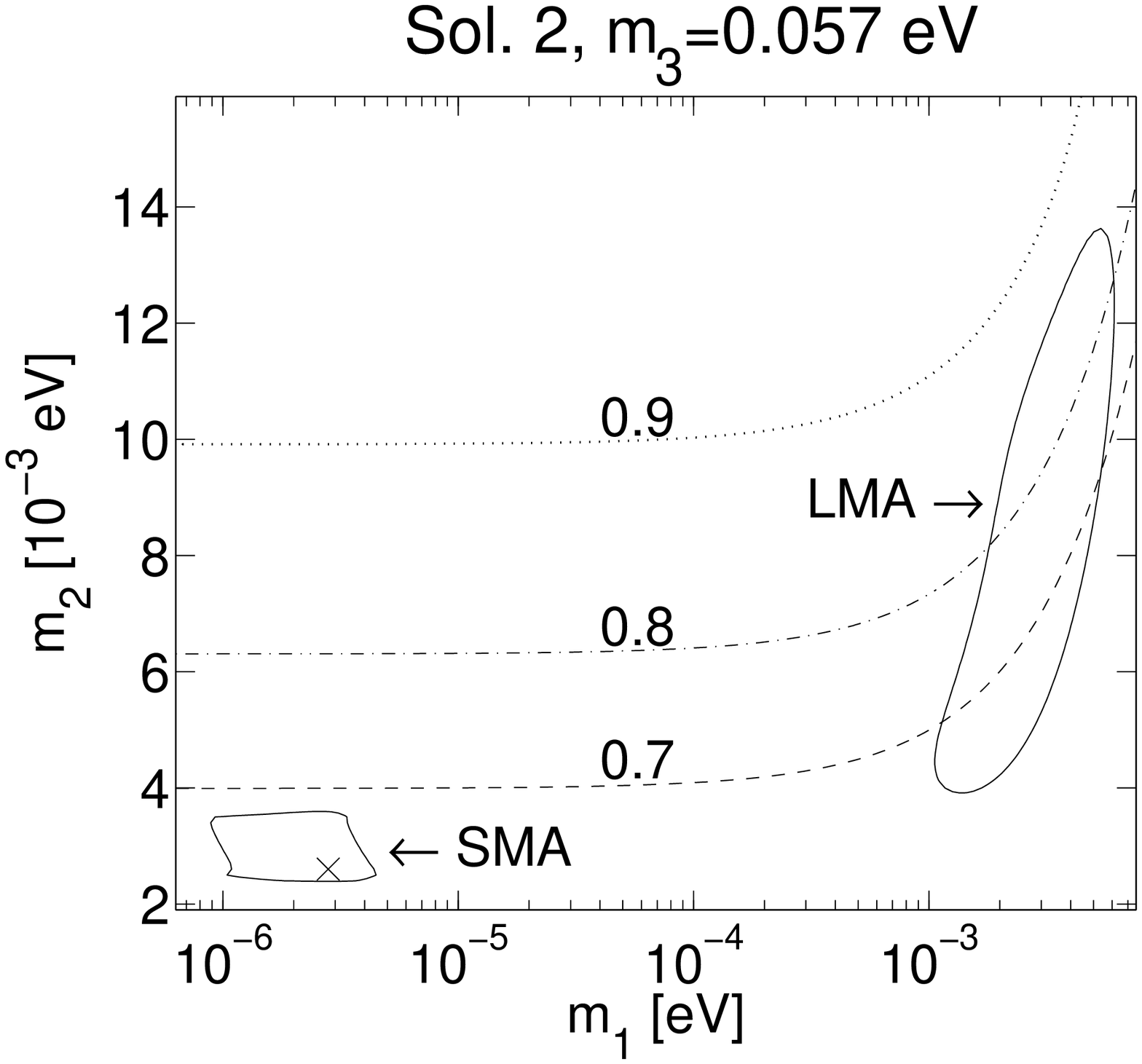}}}}  
\end{center}
\caption{The lower left and right-hand contours correspond to 
the 95\% C.L.\ for the small-mixing-angle and large-mixing-angle solutions 
of Fig.~\ref{sma-lma}, respectively.
The global minimum (``$\times$'') has $\chi_{\rm min}^2 = 1.8$. 
The dashed, dash-dotted, and dotted lines indicate values of
$4U_{\mu3}^2U_{\tau3}^2$.}
\end{figure}

In Fig.~\ref{sma-lma} (right panel) we show how the model 
compares with the data in the region known as 
the ``large-mixing-angle solution''.
In this region, which requires somewhat larger values of $m_1$,
the relevant mixing elements are very sensitive 
to changes in the mass ratios $m_1/m_3$ and $m_2/m_3$,
which are still rather small.  
For the masses shown in this figure, $U_{e3} \lsim 10^{-2}$.
The best fit (varying $m_1$ and $m_2$) has $\chi^2=3.8$ and occurs at 
$m_1=2.1 \times 10^{-3}~\mbox{eV}$, 
$\Delta m_{21}^2=3.2 \times 10^{-5}~\mbox{eV}^2$, with the constraint
$\Delta m_{31}^2 = 3.2 \times 10^{-3}~\mbox{eV}^2$.  
For Solutions~1 and 2 the atmospheric neutrino-oscillation 
amplitudes are 0.07 and 0.7, respectively.
Thus, in the LMA region, Solution~1 is unacceptable.\footnote{Updated 
results from the Gallium detectors and Super-Kamiokande
were recently presented at the XIX International Conference
on Neutrino Physics \& Astrophysics.
Those values shift the SMA solution to $m_1=2.2\times 10^{-6}~\mbox{eV}$ 
with $\Delta m_{21}^2=1.2\times 10^{-5}~\mbox{eV}^2$.
The LMA solution is relatively less changed.}

These two low-$\chi^2$ regions can be combined into a single plot,
Fig.~\ref{sma+lma}.  The solid contours enclose regions where $\chi^2 < 7.8$.
Because we have two degrees of freedom (four different kinds of detectors
and two parameters) and the global minimum has
$\chi_{\rm min}^2=1.8$, from the relation $\chi^2 = \chi^2_{\rm min} +5.99$,
these contours corresponds to 95\% C.L.
In the same plot we also show the contours for
atmospheric $\nu_\mu \leftrightarrow \nu_\tau$ amplitudes equal to 0.7, 0.8
and 0.9.  
No point in this figure gives a mass-squared difference 
$\Delta m_{31}^2$ outside of the 68\% C.L.\ region given 
by \cite{Super-K-prl}.

\subsection{Small \boldmath{$\Delta m_{32}^2$} ($m_2/m_3 \simeq 1$)}
In this case the mixing elements are far less sensitive to small changes in
the mass ratios. And for this reason the mixing elements are less adjustable
within the relevant values of the mass-squared differences.
For Solution~1 we get a best fit $\chi^2 \simeq 12$.

For Solution~2 it is unavoidable to get $U_{e2}=U_{e3}=1/\sqrt{2}$, see 
Fig.~\ref{sol1-2}. The predicted fit to the observed values corresponds to
$\chi^2\simeq 15$, which is quite poor. Small $\Delta m_{32}^2$ is also
unfavored in view of the atmospheric neutrino data.

\section{Spectrum distortion}

The Super-Kamiokande detector measures the recoil energy of the scattered
electrons coming from the process $\nu + e^- \rightarrow \nu + e^-$
\cite{spektrum}.  
For each of the solar sources, the form of the neutrino energy spectrum 
is measured in laboratories, and is independent of the total flux.  
Thus, the Super-Kamiokande measurements of the upper part of 
the $^8{\rm B}$ energy spectrum gives restrictions on the neutrino 
parameters, independent of solar models. 
If an energy dependent suppression of the flux should be established
by such a method, it would be a very strong hint for neutrino oscillations
because it would in principle not involve the uncertainties 
associated with solar models.
However, the $hep$ neutrino flux is poorly known \cite{BKS}.
This fact tends to reduce the significance of these measurements.

\begin{figure}[htb]
\refstepcounter{figure}
\label{spectrum}
\addtocounter{figure}{-1}
\begin{center}
{\epsfysize=8cm
{\mbox{\epsffile{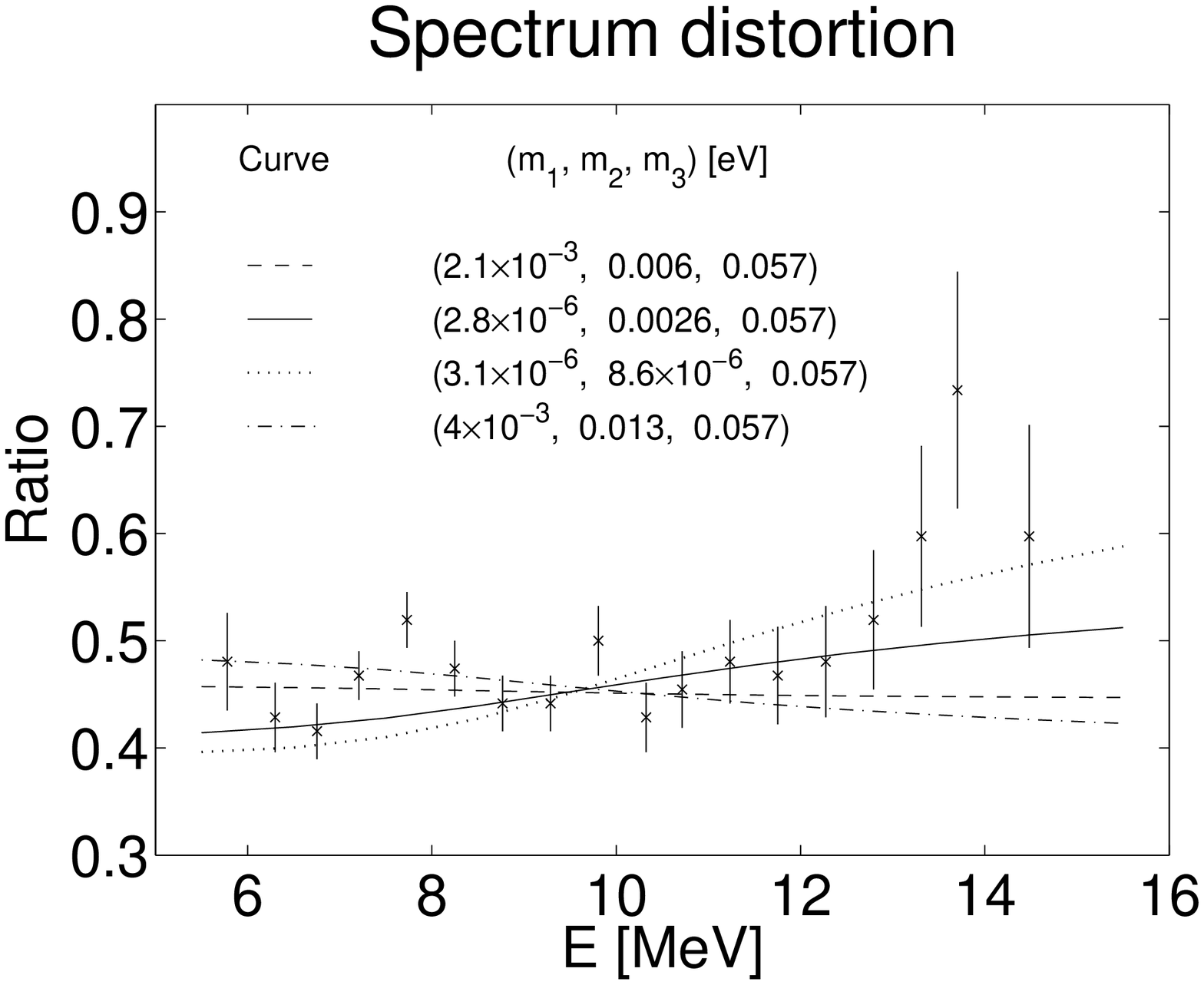}}}}  
\end{center}
\caption{Energy-dependence of recoil electrons, due to oscillations
with masses ($m_1$, $m_2$, $m_3$) as indicated in the figure:
dashed and dash-dotted for LMA-MSW,
solid for SMA-MSW, and
dotted for vacuum oscillations.
The data points show the observed attenuation w.r.t.\ the
Standard Solar Model (no oscillations) \cite{spektrum,BBP-98}.}
\end{figure}

In Fig.~\ref{spectrum} we have plotted the ratio of the measured counting
rate against the expectation in the absence of oscillations \cite{spektrum}.
If there is no neutrino flavor transition, this ratio should be a straight
horizontal line.  However, it seems to increase for $E \gtrsim 13$ MeV.  In
the same figure we have tried to reproduce the observed spectrum 
within three scenarios: the vacuum, SMA, and LMA solution.  
These are actually all acceptable, but, as concerns the spectrum, 
the vacuum and SMA solutions are favored.

\section{Conclusion}
We have extended earlier work on the application of 
the Lehmann--Newton--Wu mass matrix \cite{LNW} to neutrinos 
by investigating also the vacuum
oscillation case, and by studying in more detail the MSW solutions.
Two new fits to the solar neutrino counting rates have been found, 
beyond those of \cite{OW}, one at $m_1\sim 3\times 10^{-6}$~eV, 
with $m_2\sim 3\times 10^{-3}$~eV and $m_3\sim 0.05$~eV,
corresponding to a small mixing angle solution in the MSW interpretation,
and a vacuum-oscillation solution with $m_1$ and $m_2$ of the order
$10^{-6}$--$10^{-5}$~eV. Both of these solutions lead to electron energy 
recoil spectra that are in better agreement with the data than 
the LMA solutions,
but they are only in marginal agreement with the atmospheric neutrino
data.

For both the vacuum and MSW solutions of the solar neutrino problem, 
only hierarchical mass values gives viable fits.
In both cases, best fits are obtained when the largest mass value is
$m_3 \simeq 0.05$.  When atmospheric neutrino constraints are taken into
account, only the large-angle part of the MSW effect gives an acceptable
fit to the observed data. The result from the combined analyses of
the atmospheric and solar neutrino data will be further constrained with
the forthcoming long-baseline experiments and SNO data \cite{SNO}.
\bigskip

\leftline{\bf Acknowledgments}
\par\noindent
We are greatly indebted to Professor Tai Tsun Wu
for the most helpful discussions.  
This work was supported in part by the Research Council of Norway.

\end{document}